\documentclass[reprint,showpacs,amsmath,amssymb,aps,prl]{revtex4-1} 
\usepackage{graphicx}
\usepackage{dcolumn}
\usepackage{bm}
\usepackage{import} 
\usepackage{dsfont}
\usepackage{epstopdf}
\DeclareGraphicsExtensions{.eps,.pdf,.png,.jpg}

\begin{document}

\title{Complementary Speckle Patterns : deterministic interchange of intrinsic vortices and maxima through Scattering Media}

\author{J\'er\^ome Gateau,$^1$ Herv\'e Rigneault,$^2$ Marc Guillon$^3,*$}

\affiliation{
$^1$ Holographic Microscopy Group,Neurophotonics Laboratory, CNRS\nolinebreak[3] UMR~8250, Paris Descartes University, Sorbonne Paris Cit\'e, Paris, France\\
$^2$ Aix-Marseille University, CNRS, Centrale Marseille, Institut Fresnel UMR~7249, 13013 Marseille, France\\
$^3$ Wavefront Engineering Microscopy Group, Neurophotonics Laboratory, CNRS\nolinebreak[3] UMR~8250, Paris Descartes University, Sorbonne Paris Cit\'e, Paris, France\\
$^*$Corresponding author: marc.guillon@parisdescartes.fr
}

\date{\today}

\begin{abstract}
Intensity minima and maxima of speckle patterns obtained behind a diffuser are experimentally interchanged by applying a spiral phase delay of charge $\pm 1$ to the impinging coherent beam. This transform arises from the intuitive expectation that a tightly focused beam is so-changed into a vortex beam and {\it vice-versa}. The statistics of extrema locations and the intensity distribution of the so-generated ``complementary'' patterns are characterized by numerical simulations. It is demonstrated experimentally that the incoherent superposition of the three ``complementary speckle patterns'' yield a synthetic speckle grain size enlarged by a factor $\sqrt{3}$. A cyclic permutation of optical vortices and maxima is unexpectedly observed and discussed.
\end{abstract}

\pacs{42.25.Dd,42.25.Fx,42.30.Ms,02.40.Xx}
%

\maketitle 

Propagation of coherent waves in disordered scattering media is associated with the creation of random wavefields. The resulting inhomogeneous intensity and phase patterns are observed and studied with acoustic waves~\cite{Lopez_IEEE_83}, matter waves~\cite{Truscott_NC_11} and with electromagnetic waves from the microwave regime~\cite{Genack_nat_11} up to the X-ray regime~\cite{Stephenson_nat_91}.
Although appearing in random media, the scattering process is deterministic. Thereby, the linear relationship between the incident and scattered wavefields allows controlling the output intensity pattern~\cite{Mosk_NP_12}. However, this approach requires the tedious prior characterization of the scattering properties of the medium for each output mode~\cite{Gigan_PRL_10,Yang_NP_15}. Angular correlation properties of the scattering medium, known as the ``memory effect''~\cite{Feng_PRL_88,Stone_PRL_88}, can also be used to retrieve the image of objects hidden by a scattering medium~\cite{Labeyrie_AA_70,Gigan_NP_14,Mosk_nat_12,Psaltis_OE_14}, but necessitate a thin enough diffuser~\cite{Gigan_OE_15}. 
Therefore, scattering media are usually considered as a major obstacle for focusing and imaging. Nevertheless, the generated scattered intensity patterns (speckle) feature specific distributions and correlations~\cite{Freund_1001_correlations} that can be exploited {\it per se}, especially in optical imaging. Uncontrolled speckle patterns used as structured illuminations have been demonstrated to enhance microscopy~\cite{Sentenac_NP_12} and photoacoustic imaging~\cite{Gateau_OL_13}. Critical points such as intensity minima also enable sub-diffraction tracking of dynamic processes~\cite{Hanson_OE_06,Duncan_JBO_12}.
 Here, we consider capitalizing on both the properties of the medium and the speckle pattern through the unexplored approach : controlling deterministically intrinsic critical points of a random scattered wavefield. Full control of the scattered intensity pattern is a difficult task but predictable relative changes can be straightforward, and could modify the spatial location and type of specific speckle features.

Critical points in random scalar wavefields comprise intensity maxima but also zeros of intensity bearing a helical phase structures (vortices). These singular points appear with an equal density of topological charges $+1$ or $-1$~\cite{Berry_PRSLA_74, Berry_JPA_78, Baranova_JETP_81,Freund_PLA_95}. In free space, a vortex may typically be obtained by imprinting a spiral phase ($\rm SP$) mask to a peaked focused beam~\cite{Padgett_PO_09} and the inverse transform is obtained applying the $\rm SP$ mask of opposite charge. 
Moreover, in the context of telecommunication, it was shown that vortices could be transmitted through scattering media~\cite{Alfano_OL_16,Singh_APL_15} and across turbulent atmosphere~\cite{Tyson_JOSAA_08}.

In this Letter, we show that intensity maxima and vortices appearing behind a scattering medium can be exchanged by modulating the impinging beam with spiral phase delays of charge $\pm 1$ (${\rm SP}^{\pm 1}$). Since intensity maxima and zeros are interchanged, we shall qualify the created speckle patterns as ``complementary''. In the following, the multiple aspects of this complementarity are experimentally and numerically characterized for coherent optical waves in the light of analytical models. We demonstrate that the location exchange of vortices and intensity maxima also results in complementary spatial arrangements of speckle grains. 
Finally, intensity values at locations of critical points of complementary patterns are analyzed and discussed in the frame of a permutation algebra.

\begin{figure}[htb]
\begin{center}
\includegraphics[width=\columnwidth]{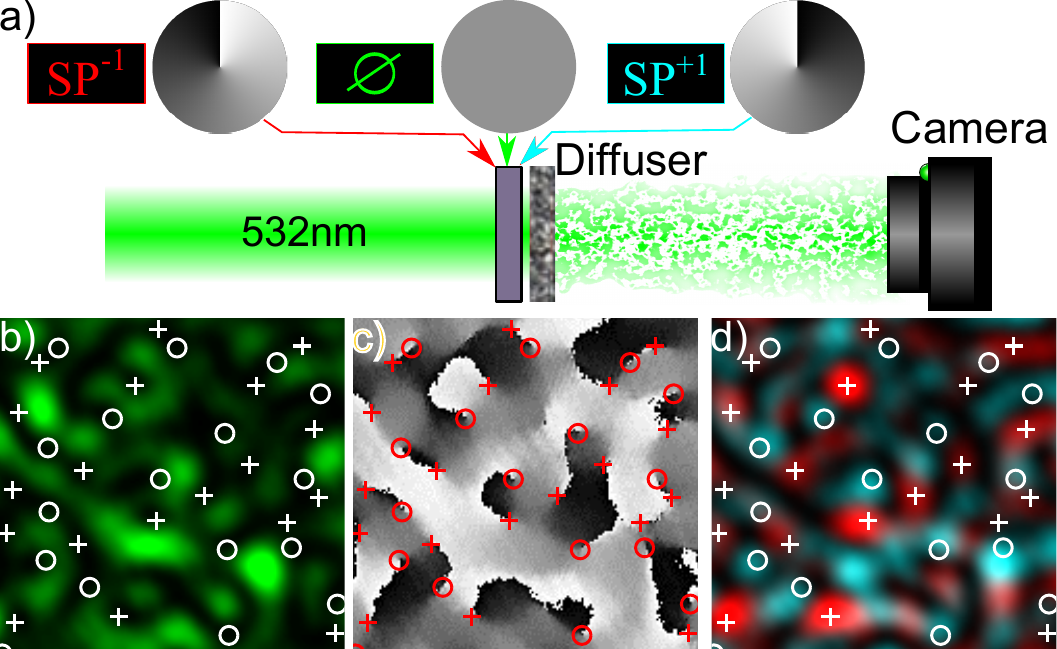}
\caption{Experimental generation of complementary speckle patterns (a): spiral phase delays of charge $-1$ and $+1$ (${\rm SP^{\mp 1}}$) are introduced by a spatial light modulator on the beam impinging onto the diffuser. Speckle patterns corresponding to each configuration are recorded on a camera. The intensity (b) and phase (c) patterns of $A_0$ were measured experimentally by phase-stepping digital holography. $+1$ and $-1$ vortices of $A_0$ are identified with $+$ and $\rm o$ symbols respectively. Intensity patterns $I_1$ (cyan) and $I_{-1}$ (red) resulting from the addition of ${\rm SP^{\pm 1}}$ are also compared to vortices of $A_0$ (d). Maxima of $I_1$ and $I_{-1}$ exhibit a high correlation with vortices of $A_0$.}
\label{fig:principle}
\end{center}
\end{figure}

The typical experimental configuration we use is shown in Fig.~\ref{fig:principle}a. A laser beam illuminates a diffuser and spiral phase delays ${\rm SP}^{\pm 1}$ may be imprinted to the wavefront with a spatial light modulator (SLM). The SP mask is placed close enough to the diffuser so that the intensity distribution of the impinging beam is minimally altered and all statistical properties of the speckle are preserved. Since energy is also preserved, intensities are normalized by their ensemble average in the following.
Wavefields and intensities associated with no phase delay and ${\rm SP}^{\pm 1}$ are notated $A_0$, $A_{\pm 1}$ and $I_0$, $I_{\pm 1}$ respectively. Maxima and $\pm 1$ vortices of $A_j$ are called $M_j$ and $V_j^{\pm}$, respectively.
The intuitive expectation that adding ${\rm SP}^{\pm 1}$ will change intensity maxima $M_0$ into zeros ($V_{\pm 1}^{\pm}$) and optical vortices $V_0^{\mp}$ into intensity maxima ($M_{\pm 1}$) is further supported by an analytical model based on a first order Taylor expansion of the wavefields under symmetry conditions on the power spectrum ~\cite{supplementary}. As demonstrated in the context of spiral phase contrast imaging~\cite{Marte_JOSAA_06,Guillon_JOSAA_14}, an estimate of $A_{\pm 1}$ in a plane transverse to the propagation axis is given by :
\begin{equation}
A_{\pm 1}^{(1)}({\bf r}) = \mp K\nabla_\perp A_0({\bf r}) \cdot \sigma_{\mp}
\label{eq:first_order}
\end{equation}
where $\sigma_+$ and $\sigma_-$ designate circular vectors ($\sigma_\pm=\left(e_x\pm ie_y\right)/\sqrt{2}$), and $K$ is a constant that depends on the power spectrum of $A_0$~\cite{supplementary}. 
At maxima of $A_0=\rho e^{i\chi}$, we have $\nabla_\perp\rho=0$. Moreover, since $\nabla_\perp \chi$ is weak due to the proximity of intensity maxima with phase saddles~\cite{Freund_OC_95}, Eq.~\eqref{eq:first_order} implies that $|A_{\pm 1}| \ll 1$. Conversely, at a vortex $V_0^{\pm}$, approximating $A_0$ by $re^{\pm i\theta}$ ($r$ and $\theta$ designating polar coordinates centered on $V_0^{\pm}$), we obtain $\nabla_\perp A_0\propto \sigma_{\pm}$, thus maximizing $|A_1|$ and minimizing $|A_{-1}|$ at $V_0^-$ and {\it vice-versa} at $V_0^+$.
An experimental illustration of $V_0^{\pm}$ switching to $M_{\mp 1}$ is shown in Fig.~\ref{fig:principle}b-d. A SLM (LCOS, X$10468$, Hamamatsu) was used to generate both the scattering phase pattern and phase masks ${\rm SP}^{\pm 1}$. The intensity $I_0$ (Fig.~\ref{fig:principle}b) and the phase of $A_0$ (Fig.~\ref{fig:principle}c) were measured by phase-stepping interferometry illuminating the parallel aligned liquid crystal SLM with a $45^\circ$ polarized beam to use the unmodulated component as a reference beam. $V_0^{\pm}$ deduced from the phase map are superposed with $I_1$ and $I_{-1}$ (Fig.~\ref{fig:principle}d) and 
validate the high correlation with $M_{\mp 1}$.
We note that some maxima in Fig.~\ref{fig:principle}d are not associated with any vortex. This point 
will be discussed later. 

\begin{figure}[htb]
\begin{center}
\includegraphics[width=\columnwidth]{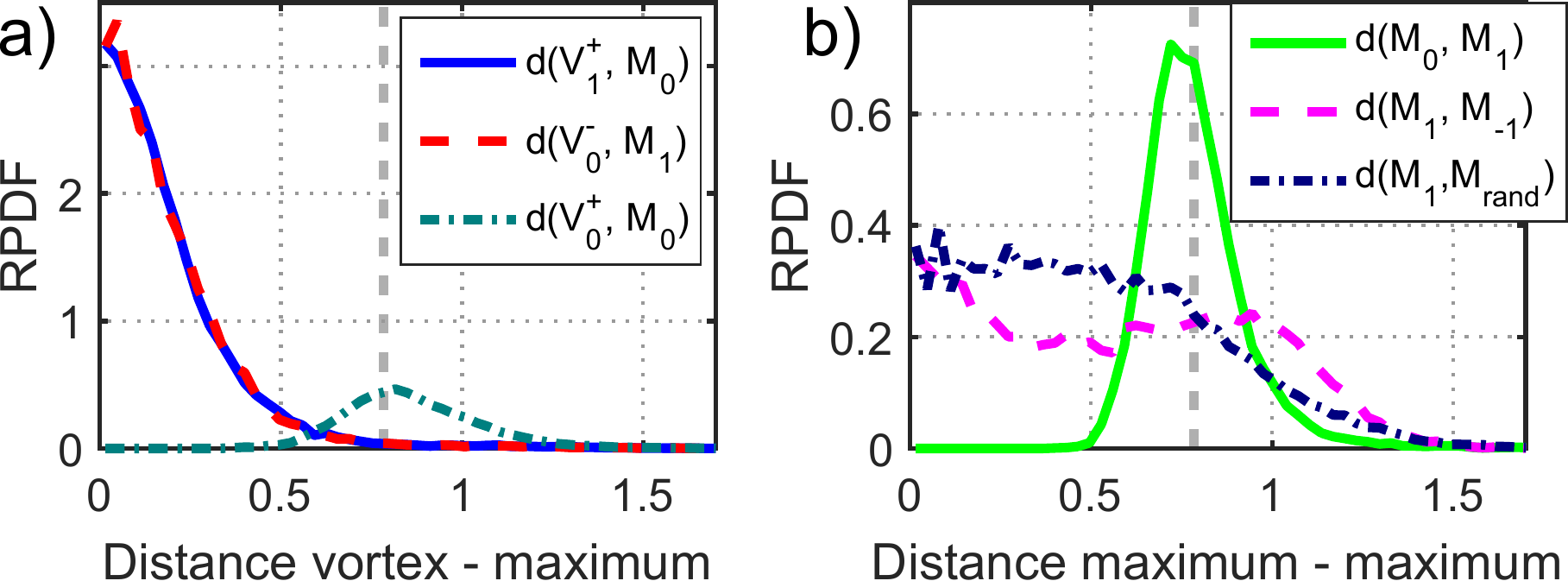}
\caption{Radial Probability Density Functions (RPDFs) of separation distances between one set of critical points and the closest points of another set obtained from numerical simulations. The abscissa unit is $\lambda/(2.{\rm NA})$. The distance between $V_1^+$ and the closest $M_0$ is notated d$\left(V_1^+, M_0\right)$. Vertical thick dashed gray lines correspond to the correlation length $l_c \approx 0.78$.}
\label{fig:distances}
\end{center}
\end{figure}

\begin{figure}[htb]
\begin{center}
\includegraphics[width=8cm]{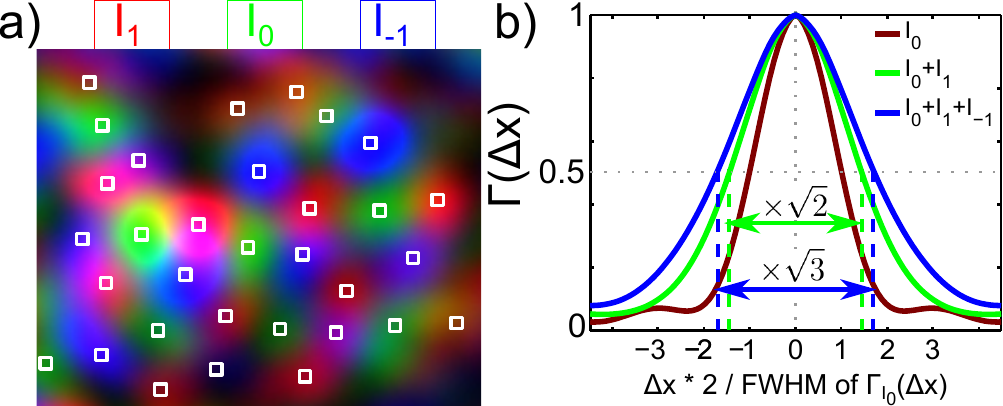}
\caption{Speckle patterns $I_0$, $I_1$ and $I_{-1}$ measured through a ground glass diffuser and overlapped, coded in saturated Red Green Blue colors (a). The locations of the maxima are marked with squares, illustrating a dense mapping of the plane. Cross-sections of the (normalized) auto-correlation functions of $I_0$, the sum $I_0+I_1$ and the sum $I_0+I_1+I_{-1}$ demonstrate the complementarity of speckle patterns (b).}
\label{fig:autocorr}
\end{center}
\end{figure}

To quantify the exchange between $M_0$ and $V_1^{+}$, we computed the radial probability density function (RPDF) of the distance between $V_1^+$ and the closest $M_0$ (notated {\it d}$\left(V_1^+, M_0\right)$) from numerical simulations~\cite{supplementary} of Gaussian random wavefields in a tranverse plane (Fig.~\ref{fig:distances}a). This RPDF(r) corresponds to the probability to find the closest $M_0$ at the distance r from a vortex $V_1^+$ per unit area. The RPDF of {\it d}$\left(V_0^-, M_1\right)$ is the same due to the symmetry of the $\rm SP$ transform (Fig.~\ref{fig:distances}a).
For comparison, the RPDF for {\it d}$\left(V_0^+, M_0\right)$ is plotted. Distances are normalized to $\lambda/(2{\rm NA})$ (where $\lambda$ is the wavelength and $\rm NA$ the numerical aperture of illumination), i.e. the full width half maximum (FWHM) of a speckle grain. The RPDF of the distance between $V_1^{+}$ and $M_0$ is tightly confined at zero distances confirming the interchange induced by the spiral phase mask, while maxima $M_0$ and vortices $V_0$ in a speckle pattern are statistically separated by a larger average distance of about one correlation length $l_c \approx 0.78$~\cite{Freund_1001_correlations}. 
The RPDF of the distance between $M_0$ and the closest $M_{\pm 1}$ (correlating with $V_0^\mp$) exhibits a exclusion distance equal to this correlation length (Fig.~\ref{fig:distances}b). The peak of this latter RPDF is centered at $l_c$, which is also the radius of the donut that would be obtained adding ${\rm SP}^{\pm 1}$ to the wavefront of a focused beam propagating in free space. Therefore, bright speckle spots surrounding maxima of $I_1$ and $I_{-1}$ will be minimally overlapping with those of $I_0$, which is verified experimentally (Fig.~\ref{fig:autocorr}a).
 For comparison, the RPDFs of the distance between $M_1$ and $M_{-1}$ as well as the distance between $M_1$ and maxima of a non-correlated speckle pattern are also shown in Fig.~\ref{fig:distances}b. Although no exclusion distance is observed between $M_1$ and $M_{-1}$, the lower RPDF of {\it d}$\left(M_1, M_{-1}\right)$ for small distances indicates a partial repulsion of bright speckle spots of $I_1$ and $I_{-1}$. The repulsions between maxima of the complementary speckle patterns reveal a new facet of the complementarity in terms of spatial arrangement of speckle grains.

Experimentally, the complementarity of spatial arrangements of bright speckle spots was first investigated with a surface scatterer (120 grit ground glass diffuser, Thorlabs). Overlaying $I_0$, $I_1$ and $I_{-1}$, speckle grains are observed to be closely packed and to create a dense mapping of the plane (Fig.~\ref{fig:autocorr}a). The characteristic length scale of the spatial fluctuation of the sum of $I_0+I_1$ is therefore observed to be larger than for a single speckle pattern (Fig.~\ref{fig:autocorr}b), contrary to the sum of two uncorrelated speckle patterns. The $\sqrt{2}$ factor on the FWHM of the autocorrelation function corresponds to a doubling of the coherence area, since speckle grains of $I_0$ and $I_1$ minimally overlap. 
Similarly, the FWHM autocorrelation function of the sum $I_0+I_1+I_{-1}$ is increased by a factor $\sqrt{3}$ (Fig.~\ref{fig:autocorr}b), corresponding to a tripling of the coherence area since $M_1$ and $M_{-1}$ partially repulse. Consequently, the induced speckle complementarity allows synthetic enlargement of speckle grains, and thus provide the possibility to uniquely build up low spatial frequencies of the intensity distribution by incoherent summation. Interestingly, the first order statistics of the sum $I_0+I_1+I_{-1}$ were verified to match these of the sum of three random speckle patterns~\cite{Goodman}, meaning that the complementarity only holds on a local scale. To go beyond surface scattering, we conducted an experiment through a $0.7~{\rm mm}$-thick slice of chicken breast, sandwiched between two glass slides. This thickness corresponds to several mean-free-paths and is even of the order of the tissue transport mean free path~\cite{Ntziachristos_NM_10}. The sample also depolarizes the incident laser beam. In this experimental configuration, $I_0+I_1+I_{-1}$ was found to yield the same enlargement of the autocorrelation function as for the surface diffuser~\cite{supplementary}. Behind the sample, the complementarity was also observed for each polarization component despite depolarization~\cite{supplementary}. The deterministic generation of complementary speckle patterns is then  robust through several scattering mean free path. We interpret that complementary speckle generation requires a minimalistic memory effect over a distance as short as a single correlation length.

\begin{figure}[htb]
\includegraphics[width=\columnwidth]{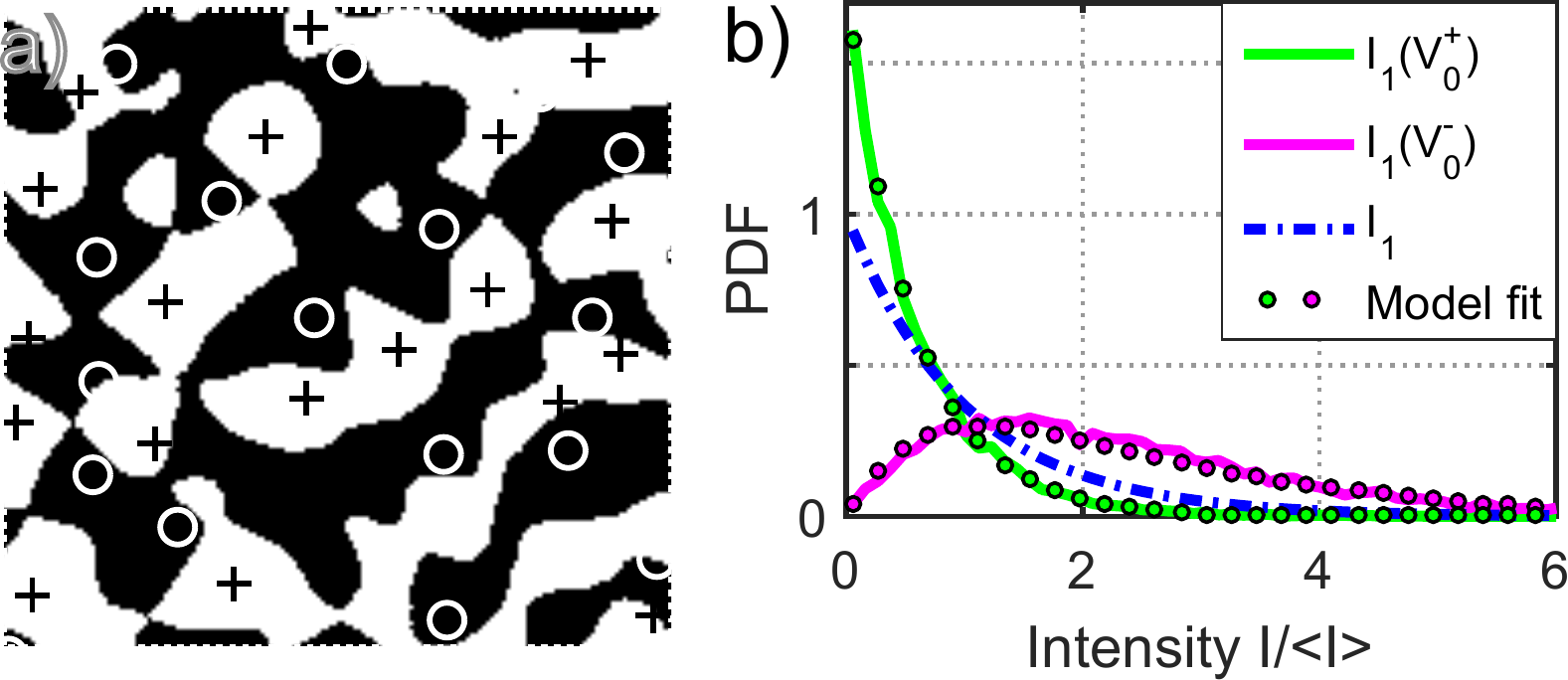}
\caption{From experimental data shown in Fig.~\ref{fig:principle}, the sign of $I_{-1}-I_1$, represented in black (negative) and white (positive), identifies the sign of vortices $V_0^{\pm}$ (a). PDFs of intensities $I_1$ at optical vortices of $A_0$ (b) deduced from numerical simulations (solid lines) and from fits of first order approximation functions. The PDFs are compared with the PDF of speckle patterns $I_1$.
}
\label{fig:pdf_vortex}
\end{figure}

In addition to spatial correlations between vortices and maxima, intensity values $I_1$ and $I_{-1}$ are expected to exhibit particular statistical properties at the locations of $V_0^{\pm}$. We derive here an analytical treatment of this statistical behavior and validate it by numerical simulation. 
For a fully developed speckle pattern exhibiting Gaussian statistics and with a power spectrum having circular symmetry, it can be shown that the joint probability density function (PDF) of complementary speckle patterns is~\cite{Goodman}:
\begin{equation}
\rho(I_0,I_1^{(1)},I_{-1}^{(1)}) = e^{ -I_0 }\times \alpha e^{-\alpha \left(I_1^{(1)}+I_{-1}^{(1)}\right) } 
\label{eq:intensity_pdf}
\end{equation}
where $I_{\pm 1}^{(1)}$ are defined according to Eq.~\eqref{eq:first_order} and $\alpha=\left<I_{\pm 1}^{(1)}\right>^{-1}$ slightly differs from $\left< I_0 \right> = 1$ due to the first order approximation. The PDF of intensities $I_{\pm 1}$ at locations of $V_0^{\pm}$ is calculated by integration of Eq.~\eqref{eq:intensity_pdf}, using the property that, at first order, the charge of the optical vortex associated with zeros of $I_0$ is given~\cite{supplementary} by the sign of $I_{-1}^{(1)}-I_1^{(1)}$ as confirmed in Fig.~\ref{fig:pdf_vortex}a, from experimental images shown in Fig.~\ref{fig:principle}. Numerical simulations of complementary Gaussian random wavefields statistically validated this property for $93\%$ of the vortices. 
In an aside, non-zero stationary points of $I_0$ (mainly maxima and saddle points) can be shown to lay on nodal lines $I_{-1}^{(1)} = I_1^{(1)}$. Again, numerical simulations of Gaussian random wavefields remarkably confirm that $97\%$ of maxima and $90\%$ of saddles lie at distances smaller than $0.25 $~\cite{supplementary}. 

The conditional PDF of $I_1^{(1)}$ at $V_0^+$ and $V_0^-$ can be calculated:
\begin{eqnarray}
\rho_{V_0^-}(I_1^{(1)}) & = & 2\alpha e^{ -\alpha I_1^{(1)} }\left(1-e^{ -\alpha I_1^{(1)} }\right) \label{eq:pdf_plus_plus_a}\\
\rho_{V_0^+}(I_1^{(1)}) & = & 2\alpha e^{-2 \alpha I_1^{(1)} } \label{eq:pdf_plus_plus_b}
\end{eqnarray}
Eq.~\eqref{eq:pdf_plus_plus_a} and~\eqref{eq:pdf_plus_plus_b} are plotted in Fig.~\ref{fig:pdf_vortex}b and were found to match PDFs from numerical simulations with fitting coefficients $\alpha=0.60$ for Eq.~\eqref{eq:pdf_plus_plus_a} and $\alpha = 0.86$ for Eq.~\eqref{eq:pdf_plus_plus_b}. The difference in the value of $\alpha$ for $+1$ and $-1$ vortices of $A_0$ is attributed here to the limit of the first order approximation. In addition, $V_0^+$ and $V_0^-$ experience asymmetric transformation in $I_1$. $V_0^-$ yield maxima of $I_1$ and the intensity at these points is statistically larger than the average intensity: $\left<I_1^{(1)}(V_0^-)\right>=3/(2\alpha)$. In contrast $V_0^+$ cannot result in $+2$-charged optical vortices of $A_1$ since such structures are unstable in Gaussian random wavefields~\cite{Freund_OC_93}. Vortex charges of the same sign can not be simply added, but still $V_0^+$ yield $I_1$-values statistically lower than the average intensity: $\left<I_{-1}^{(1)}(V_0^-)\right>=1/(2\alpha)$. The first order model is found to nicely account for the PDFs with adjustment of a single coefficient.

Integration of Eq.~\eqref{eq:intensity_pdf} at $M_0$ cannot easily result in an analytical expression~\cite{Weinrib_PRB_82}. From numerical simulations, the PDF of intensities $I_1$ at $M_0$ was found to have a negative exponential distribution with a rate parameter of the order of $2.4$. Therefore, although $M_0$ do not coincide exactly with $V_1^{\pm}$ vortices, the intensity at these points after adding ${\rm SP}^{\pm 1}$, is smaller than the average intensity. The rate parameter is nevertheless influenced by the discrepancy between the number of $M_0$ and the number of $V_1^+$, and weighted by maxima that do not turn into vortices.
To conclude, these latter results demonstrate the possibility to locally induce deterministic intensity fluctuations through scattering media. On a larger scale, the images of the intensity fluctuation across the complementary speckle patterns show an enhancement of low spatial frequency similarly as the sum $I_0+I_1+I_{-1}$~\cite{supplementary}.

\begin{figure}[htb]
\includegraphics[width=\columnwidth]{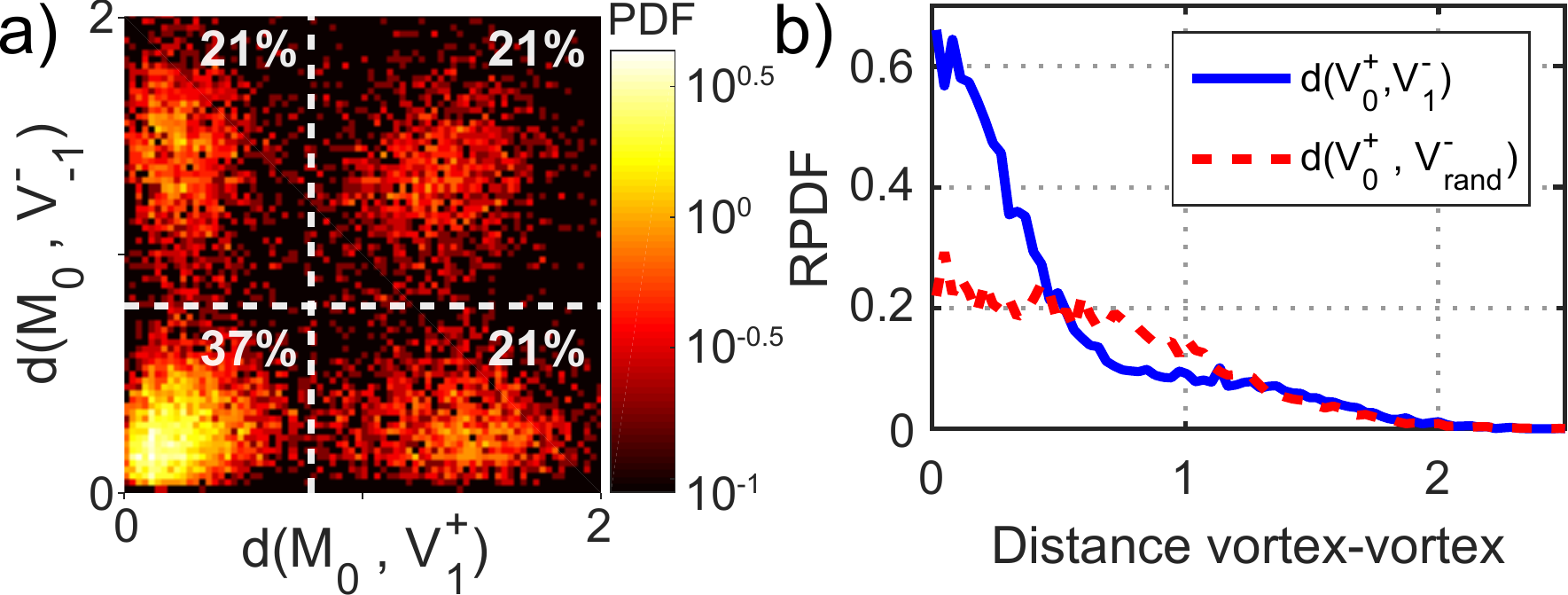}
\caption{Joint PDF of distances $d(M_0,V_1^+)$ and $d(M_0,V_1^-)$ in logarithmic color scale (a) illustrating that some maxima are not changed into vortices. The fraction of maxima populations is indicated in each quadrant. RPDF of the distance between vortices $d(V_0^+,V_1^-)$ showing correlation of the locations of these two populations of vortices.}
\label{fig:paradox}
\end{figure}

Although demonstrating the complementarity of the so-generated speckle patterns, the intuitive interpretation of the $\rm SP^{\pm 1}$ transform fail to interpret two main fundamental paradoxes. 
The first paradox arises from the different density of vortices~\cite{Berry_JPA_78,Baranova_JETP_81} and intensity maxima~\cite{Weinrib_PRB_82} in Gaussian random wavefields since these densities do not depend on the same momenta of the power spectrum. The larger density of maxima makes an one-to-one exchange impossible. In Fig.~\ref{fig:paradox}a, the joint PDF of distances $d(M_0,V_1^+)$ and $d(M_0,V_{-1}^-)$ obtained from numerical simulation is plotted for a ratio between the number of $M_0$ and the number of $V_1^+$ of $\sim 1.6$. The joint PDF exhibits two bifurcation lines at a distance of $l_c$, separating four populations of maxima $M_0$. Thus, $M_0$ is interpreted to turn into a vortex of the same charge as the $\rm SP$ mask only when the distance to the vortex is shorter than $l_c$. When a maximum $M_0$ is not changed into a vortex, we observe that the maximum distance probability for the nearest vortex is cast at $2\times l_c$. Moreover, despite vortices are less numerous, for $8\%$ of $V_0^-$ the closest $M_1$ is located beyond $l_c$ (Fig.~\ref{fig:distances}a), and thus such vortices are not transformed into maxima through a ${\rm SP}^{+1}$ mask.
The second paradox arises from the question of the zero average of the topological charge~\cite{Freund_1001_correlations}. If the topological charge of all maxima and vortices is incremented by adding a ${\rm SP}^{+1}$ mask, then, the total charge of $A_1$ would diverge as their number. Moreover, $V_0^+$ cannot possibly result in $+2$-vortices~\cite{Freund_OC_93}. 
Since we obtained from numerical simulations that $92\%$ of $V_0^-$ are changed into $M_1$ and that $92\%$ of $V_1^+$ arise from $M_0$ given a $l_c$ distance threshold, we must conclude that, in a cyclic permutation logic, most of $V_0^+$ are unexpectedly responsible for the creation of vortices $V_1^-$. We thus studied numerically the correlation in the location of $V_0^+$ and $V_1^-$. In comparison with two random distributions of vortices, the RPDF of distances $d(V_0^+,V_1^-)$ indeed exhibit a peak at small distances (Fig.~\ref{fig:paradox}b) demonstrating a  significant correlation between the locations of $V_0^+$ and $V_1^-$. The two paradoxes rise exciting fundamental questions on the rules guiding the interchange of vortices and maxima and quantification of the distances in the frame of random wavefields and their multiple correlations~\cite{Freund_1001_correlations}. They would deserve further investigation in future studies. 
The network of critical points in a random wavefield is such that even the sign of a single vortex influences the sign of all the others~\cite{Freund_PRA_94}.   

To conclude, we demonstrated the possibility to generate complementary speckle patterns through thin and thick diffusers by modulating the impinging wavefront with spiral phases. The complementarity was characterized in terms of maximum-vortex interchange and deterministic intensity fluctuation at these particular points, as well as in terms of closely packed bright speckle grains. The sum and the fluctuation of the complementary intensity patterns, on a per pixel basis, were found to synthetically enhance the low spatial frequency with as few as two speckle pattern. The limits of the one-to-one transformation of intensity extrema was discussed as well as the possibility of a cyclic permutation logic, introducing fundamental questions on the process. In spite of its fundamental complexity, the proposed complementary speckle generation bears the advantage to be easy to implement experimentally and is expected to work as long as the spiral wavefront is transmitted through the diffuser. 
Interchange of intensity maxima and vortices is at the basis of parallelized super-resolved RESOLFT microscopy~\cite{Hell_NM_13,Hell_OE_15}. Moreover, we recently demonstrated that optical vortices of speckle patterns could confine fluorescence to sub-diffraction dimensions~\cite{Pascucci_PRL_16}. Synthetic design of large speckle grain could find applications in photoacoustic imaging~\cite{Gateau_OL_13}.

\begin{acknowledgments}
The authors acknowledge Robert Kuszelewicz, Joseph Zyss and Valentina Emiliani for stimulating discussions. This work was supported by grants from the R{\' e}gion Ile-de-France, by the French-Israeli Laboratory NaBi and the Centre National de la Recherche Scientifique. 
\end{acknowledgments}

\newpage


\onecolumngrid
\begin{center}
\appendix{{\huge Complementary Speckle Patterns: Supplementary information}}
\end{center}

\twocolumngrid

\section{First order approximation of spiral transform}

$A_0$, $A_1$ and $A_{-1}$ designate the random wavefields obtained without phase mask, with a $+1$ spiral phase (${\rm SP}^{+1}$) mask and with a $-1$ spiral phase (${\rm SP}^{-1}$) mask, respectively. Here we derive $A_1$ from  $A_0$. $A_{-1}$ can then be deduced by changing the sign of the charge. Provided that $A_1$ is obtained by multiplying the wavefront at infinity by $e^{i\Theta}$ ($\Theta$ being the azimuthal coordinate in the far field), it can be expressed as a function of $A_0$ as:
\begin{eqnarray}
\label{eq:convolution}
A_1 ({\bf r}) & = & S_1 \ast A_0 \ ({\bf r})\\
    & = & \int{ S_1 ({\bf r^\prime}) A_0 ({\bf r}-{\bf r^\prime})  ds^\prime}
\end{eqnarray}
where $S_1({\bf r})={\cal F}(T)$ is the amplitude point spread function of an spiral transform given 
by the Fourier transform of the transmission coefficient $T=\Pi({\bf k})e^{i\Theta}$. $\Pi$ stands for the pupil profile and ${\bf k}$ for the spatial vector in the pupil plane. 
$T$ can then be projected on the Laguerre-Gaussian (LG) functions which form a complete basis. Since LG modes are eigenfunctions of the Fourier transform~\cite{Liu_IEEE_12}, the decomposition of $S_1$ directly arises from the decomposition of $T$. For a pupil function with circular symmetry, the decomposition of $T$ (and $S_1$) over LG functions only involves function with orbital number $l=1$:
\begin{equation}
\label{eq:pupil_summation}
\Pi({\bf k})e^{i\Theta}=\displaystyle{\sum_{n=0}^{\infty}a_n{\rm LG_n^1}}
\end{equation}
In this expression, the coefficient $a_0$ can be maximized by properly choosing the waist $w_k$ of the ${\rm LG}$ functions in the Fourier plane, thus allowing to approximate $T$ by $a_0.{\rm LG_0^1}$. For instance, for a disk-shaped pupil $\Pi$ with radius $k_{max}=2\pi \frac{NA}{\lambda}$, where $NA$ is the numerical aperture of the beam and $\lambda$ the illuminating wavelength, it can be easily shown numerically that $a_0=0.93$ for $w_k=k_{max}/2.137$, meaning that ${\rm LG_0^1}$ weights for $|a_0|^2=87\%$ of total energy in the summation in Eq.~\eqref{eq:pupil_summation}. 
For the more general case of a pupil support $\Pi$ with circular symmetry and centered energy distribution, $S_1$ is thus a function mainly described by the ${\rm LG_0^1}$ function. 
Since $S_1$ is a function peaked in the vicinity of the origin, a first order Taylor expansion for $A_0$ can be performed: 
\begin{equation}
\label{eq:taylor}
A_0({\bf r}-{\bf r^\prime}) = A_0({\bf r})-\nabla_\perp A_0({\bf r}) \cdot  {\bf r^\prime}+\mathcal{O}(\|r^\prime\|^2)
\end{equation}
Inserting this approximation in Eq.~\eqref{eq:convolution} yields for $A_1$ an expression proportional to the gradient of $A_0$:
\begin{equation}
\label{eq:first_order}
A_1^{(1)}({\bf r}) = -\nabla_\perp A_0({\bf r}) \cdot {\bf K_1}
\end{equation}
where ${\bf K_1}$ is a constant vector given by:
\begin{equation}
\label{eq:S1a}
{\bf K_1} = \int {{\bf r^\prime} S_1^\ast({\bf r^\prime}) ds^\prime}
\end{equation}
If we assume that $S_1$ has circular symmetry, then: 
\begin{equation}
\label{eq:S1b}
{\bf K_1} = K\sigma_-
\end{equation}
where $\sigma_-=({\bf e_x}-i{\bf e_y})/\sqrt{2}$ and where $K$ depends on $S_1$.

We note that the end result of this decomposition consists in approximating a spiral transform by the gradient of the field. In the Fourier domain, a spiral transform is obtained multiplying by $\Pi({\bf k})e^{i\Theta}$ while the derivative in Eq.\eqref{eq:first_order} is obtained multiplying ${\cal F}(A_0)$ by $k_x+ik_y$. 

To estimate $K$, a good criterion can be to consider that energy should be conserved when adding a phase mask at infinity. Therefore, we have $\left<|A_0|^2\right>=\left<|A_1|^2\right>$. Moreover $A_0$ varies with typical spatial scales of $\lambda/NA$. Therefore the gradient of $A_0$ is of the order of $A_0 \times NA/\lambda$, and according to Eq.~\eqref{eq:first_order}, it is estimated that ${\bf K_1}$ is of the order of $\lambda/NA$.

Equivalently, since $A_0$ can be obtained from $A_1$ by placing the complementary spiral phase mask $e^{-i\Theta}$, we may write at first order:
\begin{equation}
\label{eq:first_order_b}
A_0^{(1)}({\bf r}) = -\nabla_\perp A_1({\bf r}) \cdot {\bf K_{-1}}
\end{equation}
with:
\begin{eqnarray}
{\bf K_{-1}} & = & \int{ {\bf r^\prime} S_{-1}^*({\bf r^\prime}) ds^\prime}\\
& = & {\bf -K_1^\ast}
\end{eqnarray}
where $S_{-1}=-(S_1)^*$ is the coherent spiral point spread function corresponding to a $e^{-i\Theta}$ phase mask.

\begin{figure}[h]
\includegraphics[width=\columnwidth]{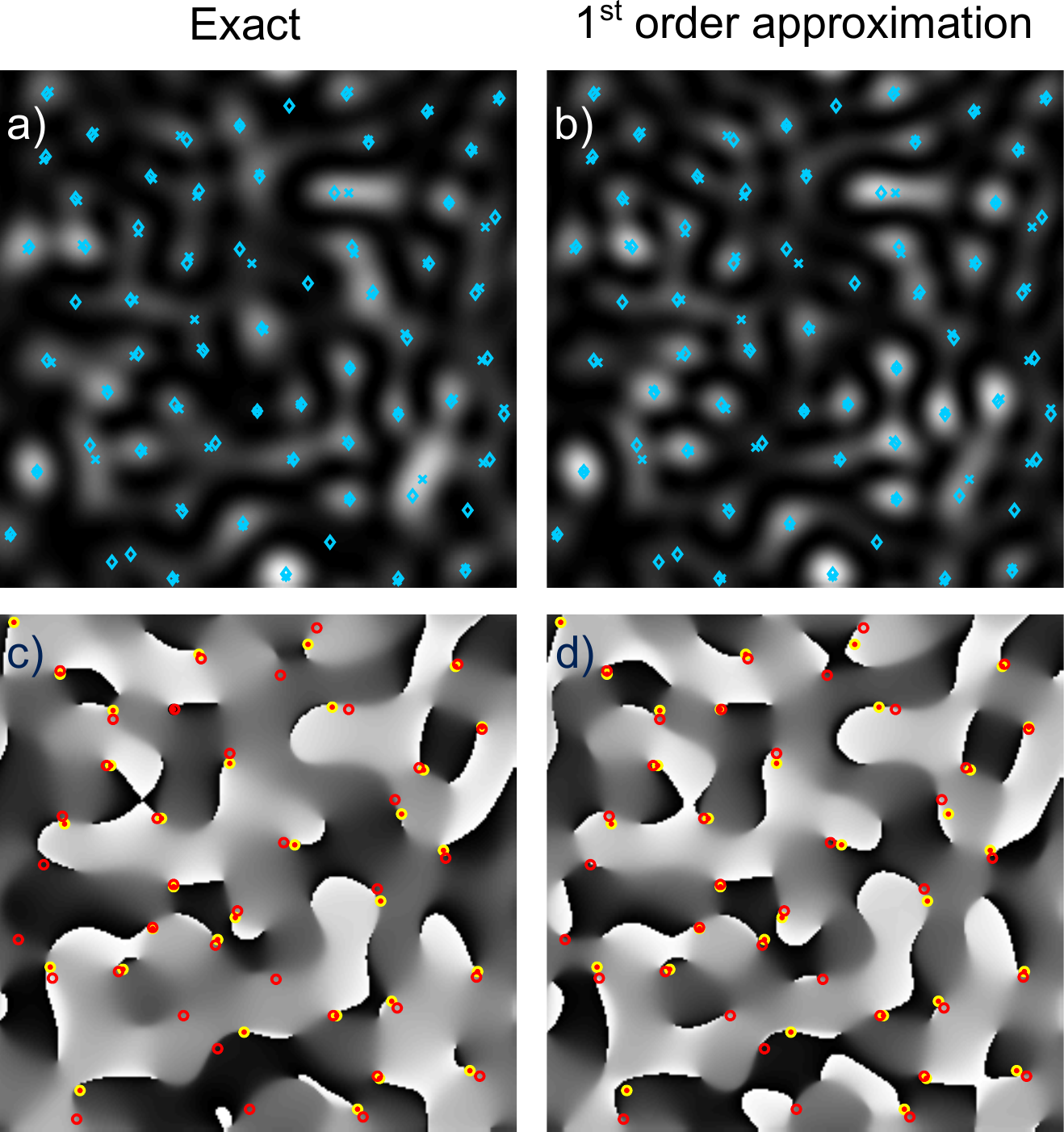}
\caption{Illustrative validation of the first order approximation given by Eq.~\eqref{eq:first_order}. Intensity (a,b) and phase (c,d) of the complementary random wavefield $A_1$. Patterns obtained from exact numerical simulation of a spiral transform (a,c) are compared with those obtained from Eq.~\eqref{eq:first_order}, the first order approximation (b,d). Intensity maxima of (a) and (b) are materialized by x-crosses and diamonds, respectively. Spiral phase singularities of charge $+1$ in (c) and (d) are materialized by plain and hollow circles, respectively.}
\label{fig:first_order}
\end{figure}

To summarize, the main hypotheses enabling the derivation of  Eq.~\eqref{eq:first_order} and Eq.~\eqref{eq:first_order_b} are that the incident beam has a circular symmetry and the Fourier transform of the transmission coefficient corresponds mainly to the Laguerre-Gaussian mode ${\rm LG_0^1}$.
 A numerical illustration of the accuracy of the first order development of Eq.~\eqref{eq:first_order} is shown in Fig.~\ref{fig:first_order}. The scalar wavefield $A_0$ was obtained here considering a top-hat circular incident beam on which random phases are imprinted, as described in the next section but on a smaller image grid. Maxima and vortices of charge $+1$ of $A^{(1)}_1$ and $A_1$ are marked and demonstrate a high spatial correlation. 

\section{Numerical simulations of Gaussian scalar random wavefields}

Numerical simulations of Gaussian scalar random wavefields were performed to have sample sizes sufficient to estimate statistical behaviors. The far-field of a uniformly-illuminated circular disk comprised of random phases was computed, without and with a ${\rm SP}^{\pm 1}$ mask, to simulate $A_0$, $A_1$ and $A_{-1}$, respectively. The transverse coherence length, i.e. the speckle grain size (FWHM), was set to $\lambda/(2.NA)= 37$ pixels where NA is the numerical aperture, and a square grid of 64 mega pixels was computed. These parameters lead to a count of critical points per generated wavefields of: $\sim 1.0 \times 10^4$ vortices of each sign, $\sim 1.58 \times 10^4$ intensity maxima and $\sim 3.29 \times 10^4$ intensity saddle points. The pixel-precise location of the points were determined using the topology of each pixel neighborhood. The radial probability function of the distance between pairs of points was normalized to account for the discretization of the images in square pixels.

\section{Correlation between critical points of $I_0$ and $I_{-1}-I_1$ }

Writing $A_0=\xi+i\eta$ (with $\xi$ and $\eta$ the real and imaginary part of $A_0$, respectively), the charge of an optical vortex is given by the sign of the vorticity vector $\omega$ projected on the propagation axis \cite{Berry_PRSLA_2000}: 
\begin{equation}
\label{eq:omega_def}
\omega\cdot {\bf e_z}=\partial_x\xi \partial_y\eta-\partial_y\xi \partial_x\eta
\end{equation}

From Eq.~\eqref{eq:first_order}, it is straightforward to show that :
\begin{equation}
\label{eq:omega}
I_{-1}^{(1)}-I_1^{(1)} = \frac{K^2}{2}\omega\cdot{\bf e_z}
\end{equation}

\begin{figure}[h]
\includegraphics[width=\columnwidth]{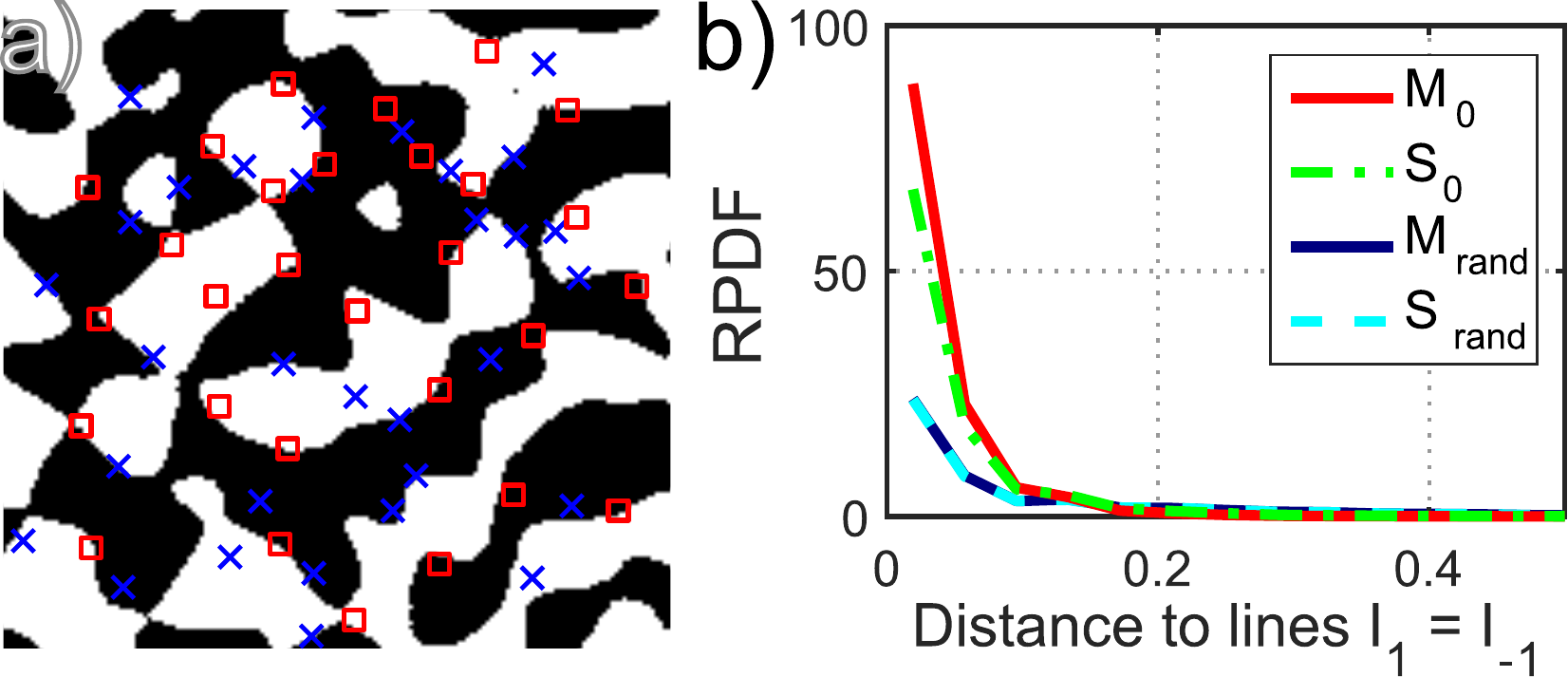}
\caption{Correlation between critical points of $I_0$ and $I_{-1}-I_1$. From experimental data shown in Fig.1, $I_{-1}-I_1$ is shown as a binary map where positive values are in white and negative ones are in black. The nodal lines $I_{-1}-I_1=0$ (borders of white areas) coincide with maxima $M_0$ (square) and saddle points $S_0$ (cross) of $I_0$ (a). From numerical simulations, the radial probability density functions (RPDFs) of the distance from the maxima $M_0$ and saddle points $S_0$ to the nodal lines (b). For comparison, the RPDFs of the distance from the maxima $M_{rand}$ and saddle points $S_{rand}$ of a non-correlated speckle pattern to the nodal lines $I_{-1}-I_1=0$ is displayed. The RPDFs show that $M_0$ and $S_0$ are closer to the nodal lines. The abscissa unit is $\lambda/(2.{\rm NA})$.}
\label{fig:distance_lignes}
\end{figure}

Another remarkable correlation can be deduced from Eq.~\eqref{eq:omega} between $I_{-1}-I_1$ and non-zero stationary points of $I_0$. At these latter locations, we have $\nabla I_0={\bf 0}$. Excluding zero solutions from this vector equation yields $\partial_x\xi \partial_y\eta-\partial_y\xi \partial_x\eta=0$, meaning that these points lie on nodal lines (or surface in three dimensions) of $\omega\cdot {\bf e_z}\propto I_{-1}^{(1)}-I_1^{(1)}$.
Fig.~\ref{fig:distance_lignes}a illustrates form experimental data that $M_0$ and $S_0$ are in the close vicinity of the nodal lines $I_{-1}-I_1=0$. Numerical simulations (Fig.~\ref{fig:distance_lignes}b) demonstrate that maxima $M_0$ and saddle points $S_0$ of $I_0$ statistically lie closer to nodal lines than maxima and saddle points of a non-correlated speckle pattern. 

\section{Complementary speckle generation through a ${\rm 700\mu m}$-thick slice of chicken breast}

\begin{figure}[t]
\includegraphics[width=\columnwidth]{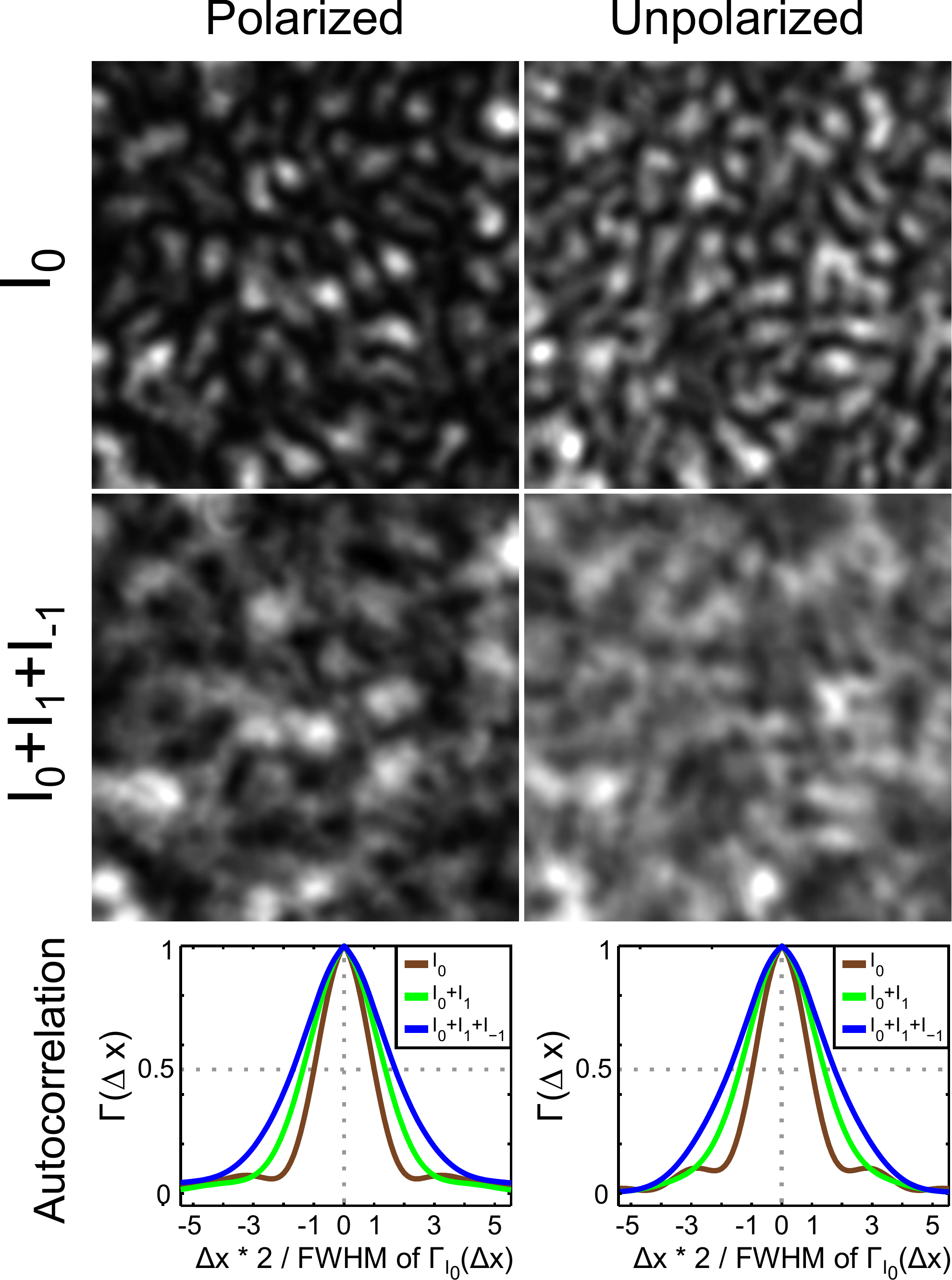}
\caption{Speckle generated behind a chicken breast slice of $700\mu m$ illuminated with a polarized laser beam, without (right column) and when adding a polarizer behind the slice (left column). Single speckle pattern (top row) and numerical incoherent addition of the three complementary speckle patterns (middle row). The autocorrelation functions of these speckle patterns are plotted (bottom row) and demonstrate that each polarization component exhibit exactly the same enlargement of the speckle grain size. This result demonstrate that complementary speckle generation also works for the polarization component arising from depolarization by the tissue.}
\label{fig:chicken}
\end{figure}

Fig.~\ref{fig:chicken} illustrates speckle patterns recorded behind a ${\rm 700\mu m}$-thick slice of chicken breast tissue. Speckle patterns presented in the right column result from the depolarized laser beam. Therefore, the obtained speckle patterns can be interpreted as the incoherent summation of two non-correlated speckle patterns, each corresponding to an orthogonal transverse polarization respectively. Despite the depolarization, the autocorrelation functions of the sums $I_0 + I_1 $ and $I_0 + I_1 + I_{-1}$ exhibit an enlargement of the FWHM by factors $\sqrt{2}$ and $\sqrt{3}$, respectively. Speckle patterns presented in the left column correspond to scattered wavefields linearly polarized (arbitrary orientation) behind the scattering medium, and are identical to speckle patterns obtained through a ground glass diffuser.
Gray scale images of the sums $I_0 + I_1 + I_{-1}$ (second row) enable to visually assess the synthetic enlargement of the correlation area.

\section{Sum and fluctuation of complementary speckle patterns enhance low spatial frequencies}

\begin{figure*}[b]
\centering \includegraphics[width=0.8\paperwidth]{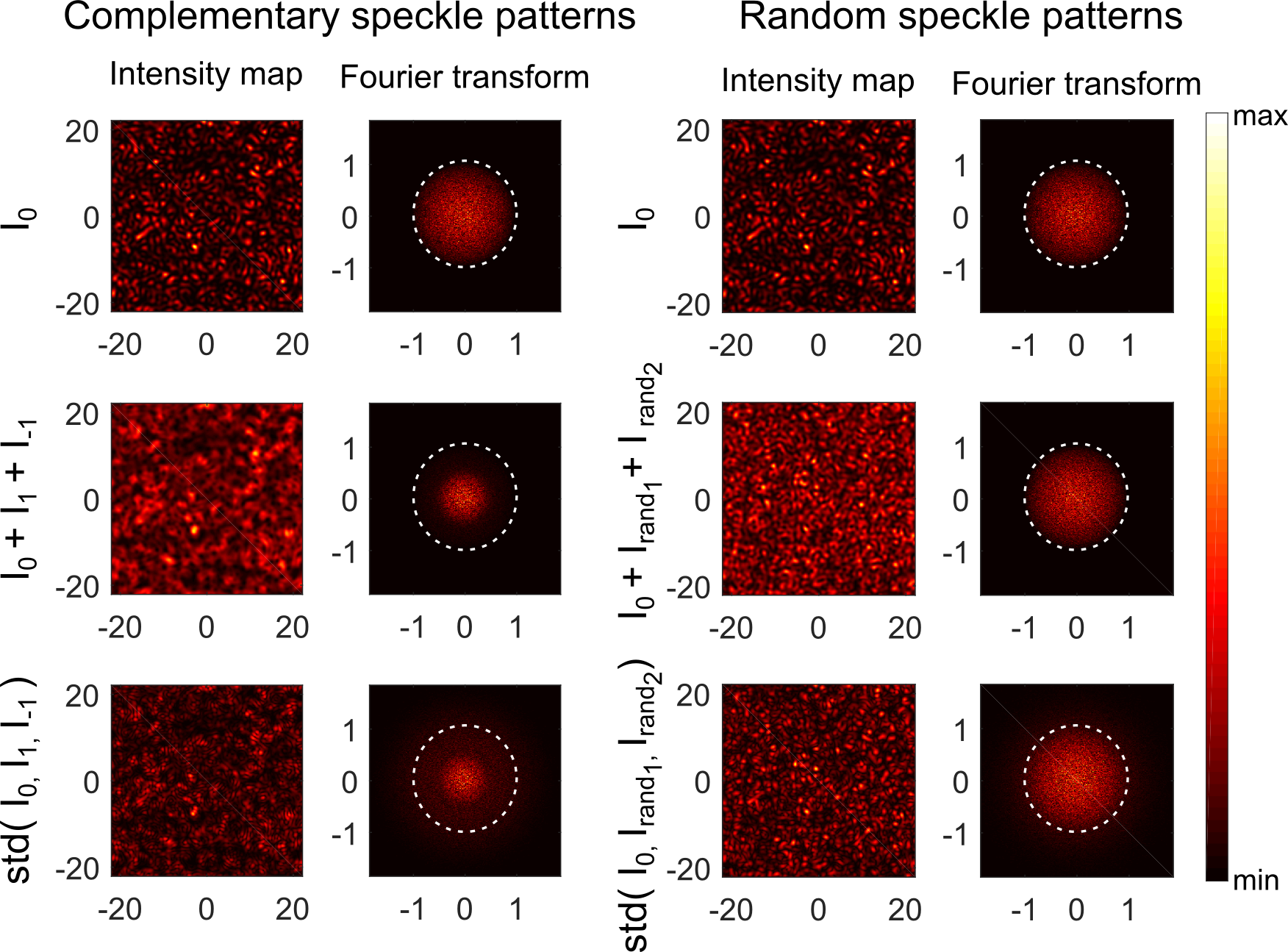}
\caption{From numerical simulation, intensity maps (first column) and magnitude of the Fourier transform (second column) of $I_0$, the sum $I_0+I_1+I_{-1}$, and the per-pixel standard deviation of $I_0$, $I_1$ and $I_{-1}$, respectively. For comparison, intensity maps (third column) and magnitude of the Fourier transform (forth column) are displayed for the sum and per-pixel standard deviation of three non-correlated speckle patterns. For the sake of visibility, the mean value was substracted to the intensity maps before computing the Fourier transform. The length unit is $\lambda / (2{\rm NA})$, and the frequency unit is $(2{\rm NA}) / \lambda$. Dashed white circles mark the unit circle.}
\label{fig:FFT}
\end{figure*}

Fig.~\ref{fig:FFT} compares the sum of complementary speckle patterns $I_0+I_1+I_{-1}$ and the per-pixel standard deviation of $I_0$, $I_1$ and $I_{-1}$ obtained from numerical simulation (first column), as well as the magnitude of their 2D Fourier transform (second column), to the sum and per-pixel standard deviation of three non-correlated speckle patterns. 
The Fourier transform of the sum of the complementary speckle patterns shows an enhancement of the low spatial frequencies in comparison with a single speckle pattern. This result corresponds to the enlargement of the auto-correlation function observed experimentally, and was therefore expected. 
The standard deviation quantifies the per-pixel fluctuation across $I_0$, $I_1$ and $I_{-1}$. The exchange between intensity maxima and zeros and the finite correlation length in speckle patterns lead to fluctuations similar on a local scale, thus enhancing the low spatial frequency components in the standard deviation image.

While the sum of non-correlated speckle patterns leads to the same spectral support as a single speckle pattern (bounded to the circle of unity), the standard deviation contain higher spatial frequencies. For complementary speckle patterns, a similar enlargement of the support is observed, but it is dominated by an building up of low spatial frequency. 

The speckle images illustrate the synthetic reinforcement of low spacial frequency for the sum and fluctuation of complementary speckle pattens in comparison to random speckle patterns. 

%

%

\end{document}